# Biophysical characterization of membrane phase transition profiles for the discrimination of Outer Membrane Vesicles (OMVs) from *Escherichia coli* grown at different temperatures

*Angelo Sarra,[1,§] Antonella Celluzzi,[2,§] Stefania Paola Bruno,[2] Caterina Ricci,[3] Simona Sennato,[4] Maria Grazia Ortore,[3] Stefano Casciardi,[5] Paolo Postorino,[4] Federico Bordi[4,#,*] and Andrea Masotti[2,#,*]*

1) Department of Science, University of Roma Tre, Via della Vasca Navale, 00146, Rome, Italy
2) Children's Hospital Bambino Gesù-IRCCS, Research Laboratories, V.le di San Paolo 15, 00146, Rome – Italy
3) Department of Life and Environmental Sciences, Marche Polytechnic University, Via brecce bianche, Ancona, I-60131, Italy
4) CNR-ISC UOS Sapienza and Department of Physics, Sapienza University of Rome, Pz.le Aldo Moro, 00185, Rome, Italy;
5) National Institute for Insurance against Accidents at Work (INAIL), Department of Occupational and Environmental Medicine, Epidemiology and Hygiene, 00078 Monte Porzio Catone, Italy.

§ These two authors equally contributed to the work

# These two authors equally contributed to the work

**Corresponding authors:**

**Andrea Masotti,** Bambino Gesù Children's Hospital-IRCCS, Research Laboratories, 00146 Rome, Italy. E-mail: andrea.masotti@opbg.net. Tel: +39 06 6859 2650.

**Federico Bordi**, Department of Physics, Sapienza University of Rome, Pz.le Aldo Moro, 00185, Rome, Italy. E-mail: federico.bordi@roma1.infn.it. Tel: +39 06 4991 3503.

# ABSTRACT

Dynamic Light Scattering (DLS), Small Angle X-ray Scattering (SAXS) and Transmission Electron Microscopy (TEM) are physical techniques widely employed to characterize the morphology and the structure of vesicles such as liposomes or human extracellular vesicles (exosomes). Bacterial extracellular vesicles are similar in size to human exosomes, although their function and membrane properties have not been elucidated in such detail as in the case of exosomes. Here, we applied the above cited techniques, in synergy with the thermodynamic characterization of the vesicles lipid membrane using a turbidimetric technique to the study of vesicles produced by Gram-negative bacteria (Outer Membrane Vesicles, OMVs) grown at different temperatures. This study demonstrated that our combined approach is useful to discriminate vesicles of different origin or coming from bacteria cultured under different experimental conditions. We envisage that in a near future the techniques employed in our work will be further implemented to discriminate complex mixtures of bacterial vesicles, thus showing great promises for biomedical or diagnostic applications.

# INTRODUCTION

In all domains of life Eukarya, Archaea and Bacteria produce and release membrane vesicles for reasons that are still not completely understood (Deatherage and Cookson, 2012). In humans, many cells such as dendritic cells, lymphocytes, and tumor cells actively release (i.e., by exocytosis) small (~30-100 nm of diameter) membrane vesicles, referred as exosomes, into biofluids (i.e., plasma/serum, urine, cerebrospinal fluid and saliva). These vesicles are powerful cell-to-cell messengers as they transfer lipids, proteins, DNA and ribonucleic acids (i.e., mRNA, microRNA, lncRNA and other RNA species) between cells (Simpson et al., 2009; Chen et al., 2012; Valadi et al., 2007). In the last few years, exosomes and their inner content have been also exploited as innovative and effective biomarkers for the diagnosis of many different diseases (i.e., tumors, obesity, gastrointestinal disorders, fibromyalgia, etc.) (Logozzi et al., 2009; Taylor and Gercel-Taylor, 2008; Masotti et al., 2017; Felli, Baldassarre and Masotti, 2017).

Similarly to human cells, either Gram-negative and Gram-positive bacteria produce extracellular vesicles, referred as Outer-membrane vesicles (OMVs) and membrane vesicles (MVs), respectively. The production and release of vesicles by bacteria is a natural process and it is necessary for inter-species (bacteria-bacteria) and inter-kingdom (bacteria-host) interactions (Leitão and Enguita, 2016).

One of the main structural differences between Gram-positive and Gram-negative bacteria relies on the composition of their cell envelope. In both cases, the bacterial envelope comprises the plasma membrane and a layer of peptidoglycans. However, in the case of Gram-negative bacteria an additional layer (i.e., the outer membrane) is also observed. The space between the cytoplasmic membrane and the outer membrane is the periplasmic space or periplasm. The outer membrane is made by negatively charged phospholipids and lipopolysaccharides that confer to the Gram-negative wall an overall negative charge. Owing to the presence of these lipopolysaccharides on the outer membranes, many gram-negative bacteria are often pathogenic.

Gram-negative bacteria (e.g. *Escherichia coli*, *Pseudomonas aeruginosa*, etc.) spontaneously secrete OMVs, small (20-100 nm of diameter) spherical bi-layered vesicles, delivered in a variety of environments, including planktonic cultures, fresh and salt water, biofilms, inside eukaryotic cells and within mammalian hosts (Beveridge, 1999; Beveridge et al., 1997; Biller et al., 2014; Hellman et al., 2000). Gram positive bacteria as well, such as *Staphylococcus aureus*, *Bacillus subtilis*, *Bacillus anthracis*, *Streptomyces coelicolor*, *Listeria monocytogenes*, *Clostridium perfringens*, *Streptococcus mutants*, and *Streptococcus pneumoniae* spontaneously produce MVs (Lee et al., 2009; Rivera et al., 2010; Schrempf et al., 2011; Lee et al., 2013; Jiang et al., 2014; Liao et al., 2014).

Recently, it has been observed that also non-pathogenic bacteria (i.e., probiotics), such as Lactobacilli, release extracellular vesicles. Given the importance and the positive impact that probiotic effects have on human health, the study of Lactobacilli MVs may be an interesting opportunity for various applications, from vaccines to therapeutic delivery (Dean et al., 2017; van der Pol, Stork and van der Ley, 2015). In this context, a detailed physicochemical characterization of MVs from *Lactobacillus acidophilus* ATCC 53544, *Lactobacillus casei* ATCC 393, and *Lactobacillus reuteri* ATCC 23272, was recently reported (Dean et al., 2019; Grande et al., 2017). Among the Lactobacilli, the *Lactobacillus rhamnosus GG* (LGG), is emerging as an important probiotic strain due to its validated effects in both treating and preventing some gastro-intestinal diseases (Wolvers et al., 2010). Behzadi and colleagues revealed the cytotoxic and inhibitor role of LGG-derived extracellular vesicles on hepatic cancer cells, but the characterization of these vesicles has not still been reported (Behzadi, Mahmoodzadeh Hosseini and Imani Fooladi, 2017). Moreover, owing to the recent employ of OMVs and MVs as adjuvants in vaccines (van der Pol, Stork and van der Ley, 2015; Bottero et al., 2016) or as novel vaccines platform (Wang et al., 2018) to regulate host activity processes (Grandi et al., 2017; Chen et al., 2017) or other pathogenic processes (Ellis and Kuehn, 2010), an in depth biophysical characterization (i.e., identification, discrimination and quantification) of these types of vesicles is highly desired for biomedical applications.

Among the most employed biophysical techniques available to characterize such small particles in aqueous suspension, Dynamic Light Scattering (DLS) represents one of the most versatile ones (see for example (Berne and Pecora, 1976)). DLS is generally used to determine vesicles size but at the same time it provides information about the reorganization of lipid molecules of the membrane bilayer with a turbidimetric method (Michel et al., 2006). Actually, the intensity of light scattered at a fixed angle by particles in a suspension depends on their size and on their optical properties (hence on the structure and on the refractive index of the components). By measuring the time-averaged intensity of scattered light as a function of the temperature, this experiment allows the determination of thermotropic lipid phase transitions (with characteristic transition temperatures Tc), related to the lipid organization and their remodeling within the bilayer, which affects the membrane optical properties (refractive index) (Aleandri et al., 2012). Consequently, if different thermotropic behaviours feature different kind of vesicles, DLS experimental set-up can be used as a tool to discriminate them, too. In fact, it was previously known that *E. coli* outer membranes show a phase transition due to conformational changes of their lipid components (Trauble and Overath, 1973). Lipid composition modifications have been also observed in their outer membranes when the environmental conditions are changed (i.e. at growing temperatures) (Morein et al., 1996; Mika et al., 2016; Marr and Ingraham, 1962). A similar behavior can be expected also for OMVs produced by *E. coli* as well.

Therefore, aimed at providing a useful tool to characterize better OMVs and MVs, we propose a solid set of experimental physical techniques, including this turbidimetric approach that is able to discriminate extracellular vesicles originating from bacteria grown at different environmental conditions.

## MATERIALS AND METHODS

### Bacterial strains and culturing conditions

*Escherichia coli* (ATCC 8739) was cultured at three different temperatures: 20°C, 27°C and 37°C in essential M9 microbial growth medium consisting in $Na_2HPO_4$ (6.8 g/L), $KH_2PO_4$ (3 g/L), $NH_4Cl$ (1 g/L) and NaCl (0.5 g/L). To M9 medium, a solution containing D-glucose (4 g/L), $MgSO_4$ (241 mg/L) and $CaCl_2·2H_2O$ (15 mg/L) was added and the resulting solution was adjusted to pH=7. Bacteria were cultured overnight until the culture reached an $OD_{600}$ of approximately 1. *Lactobacillus rhamnosus LGG* (ATCC 53103) (Dicoflor 60 ®, DICOFARM) was cultured overnight in a De Man, Rogosa and Sharpe medium (SigmaAldrich, MRS broth, 51 g/L) at 37°C in anaerobic conditions up to an $OD_{600}$ of approximately 2. Aliquots (5 ml) of the bacterial cultures (both *E. coli* and *L. rhamnosus*) were retained for further analysis and to be used as controls.

To isolate either OMVs or MVs, bacteria were removed from culture media by centrifugation (Beckman Avanti J-25 centrifuge, JA-10 rotor, 6000 × *g*, 15 min), and the supernatant was filtered through a 0.45 µm filter unit (Sartorius). The supernatant, which contains vesicles, was concentrated by ultrafiltration (Vivaflow 200, Sartorius) up to small volume (50 mL), then filtered through a 0.45 µm filter.

### Isolation of bacterial extracellular vesicles

Bacterial vesicles (OMVs or MVs) were separated by ultracentrifugation (107,000g for 3h at 4°C) (Optima™ XPN-100 Ultracentrifuge, Beckman Coulter, SW 40 Ti-rotor) of the ultrafiltered solution obtained previously. The pellet was washed once in sterile phosphate buffered saline (Dulbecco's Phosphate-Buffered Saline – PBS) and suspended in the same buffer.

At the end of the isolation protocol, a drop of the purified solution was cultured on agar plates at 37°C for one day to exclude the presence of residual bacteria.

The suspension obtained with this method contains a concentrated amount of vesicles (OMVs or MVs) that have been characterized by dynamic light scattering and turbidimetric analyses.

**Dynamic Light Scattering and turbidimetric analyses**

DLS experimental set-up has been employed to provide vesicles size and to characterize the bilayer organization because of its sensitivity to the sample refractive index (Berne and Pecora, 1976). In this case, we use the term 'turbidimetric' measurements for this unconventional way of using a light scattering apparatus, although this term is not strictly appropriate, because we observe changes in the refractive index of the lipid membranes, concerning the scattered light and not the transmitted light, as in conventional turbidimetric measurements.

DLS and turbidimetric measurements were performed employing a MALVERN Nano Zetasizer apparatus equipped with a 5 mW HeNe laser (Malvern Instruments LTD, UK). This system uses backscatter detection, i.e. the scattered light is collected at 173°. To obtain the size distribution, the measured autocorrelation functions were analyzed using the CONTIN algorithm (Provencher, 1982). Decay times are used to determine the distribution of the diffusion coefficients $D_0$ of the particles, which in turn can be converted in a distribution of apparent hydrodynamic diameter, $D_h$, using the Stokes-Einstein relationship $D_h = k_B T/3\pi\eta D_0$, where $k_B$ is the Boltzmann constant, T the absolute temperature and η the solvent viscosity (Berne and Pecora, 1976). The values of the radii shown in this work correspond to the average values on several measurements and are obtained from intensity weighted distributions (Provencher, 1982; De Vos, Deriemaeker and Finsy, 1996). The thermal protocol used for both OMVs, MVs and bacteria in DLS measurements consists of an ascending ramp from 10°C to 45°C with temperature step of 1°C. At each step, samples were left to thermalize at the target temperature for few minutes. In particular, for samples grown at 37°C, 27°C and 20°C the thermalization times were 360 s, 720 s and 1080 s, respectively.

For the turbidimetric determination of vesicles' membranes thermotropic behavior, we measured the mean count rate of scattered photons (i.e. the time-averaged intensity of the scattered light), I, versus temperature, T, and fitted the data to a Boltzmann sigmoidal curve:

$$I = I_0 + \frac{I_1 - I_0}{1 + exp\frac{T-T_0}{\Delta T}}$$  (Equation 1)

where the fitting parameters are $I_0$ and $I_1$, the minimum and maximum intensity respectively, the transition temperature $T_0$ and the so called 'slope' $\Delta T$ that describes the steepness of the curve. *ΔT* roughly measures the transition width.

**Small Angle X-ray Scattering**

SAXS experiments were performed at the Austrian SAXS beamline of the Elettra Synchrotron in Trieste, Italy (Amenitsch, Bernstorff and Laggner, 1995). SAXS images were recorded with a 2D pixel detector Pilatus3 1M spanning the Q-range between 0.1 and 6 nm$^{-1}$, with Q=4πsin(2Θ)/λ, where 2Θ is the scattering angle and λ = 0.0995 nm the X-ray wavelength. The image conversion to one-dimensional SAXS pattern was performed with FIT2D (Hammersley, 2004). Detector calibration was performed with silver behenate powder (d-spacing =0.05838 nm). Samples were held in a 1 mm glass capillary (Hilgenberg, Malsfeld, Germany). Experiments were carried out between 10 and 45°C, by using the same heating rate already adopted in DLS experiments. Each measurement was performed for 1 s and followed by a dead time of at least 10 s in order to avoid radiation damage. We obtained a set of SAXS curves resulting from the average of several overlapping SAXS spectra obtained at about the same temperature (±1°C), in order to analyze data with an improved signal-to-noise ratio. SAXS data were fitted according to a core-shell model (Berndt, Pedersen and Richtering, 2005) by GENFIT software (Spinozzi et al., 2014), and taking into account the vesicles polydispersion, in agreement with the DLS results. The fitting parameters are OMVs radius R and thickness d, as well as the electron densities of the core and of the shell, ϱi and ϱe, respectively.

**Transmission Electron Microscopy**

Transmission electron microscopy (TEM) was used for the morphological characterization of OMVs and MVs. TEM images were performed at Department of Occupational and Environmental Medicine, Epidemiology and Hygiene of INAIL-Research in Rome. Samples for TEM measurements were prepared at room temperature by depositing 20 µl of vesicles suspensions on a 300-mesh copper grid for electron microscopy covered by a thin amorphous carbon film. A negative staining was realized by addition of 10 μl of 2 % aqueous phosphotungstic acid (PTA) solution (pH-adjusted to 7.3 using 1 N NaOH). Measurements were carried out by using a FEI TECNAI 12 G2 Twin (FEI Company, Hillsboro, OR, USA), operating at 120 kV and equipped with an electron energy loss filter (Biofilter, Gatan Inc, Pleasanton, CA, USA) and a slow-scan charge-coupled device camera (794 IF, Gatan Inc, Pleasanton, CA, USA).

**Scattering cross section simulations**

To better understand the behaviour of phase transitions in OMVs and bacteria, Mie scattering simulations were performed using the MiePlot 4.6 simulator by P.Laven (Laven, 2003). This simulator allows to calculate the scattered intensity as a function of scattering angle for inhomogeneous spheres where the refractive index is a function of the radius. For simplicity sake, both OMVs and bacteria were described as multiple-shelled spheres. In all calculations, we assumed a wavelength of 633 nm for the incident light and a scattering angle of 173°. For the refractive index of the aqueous core (of both vesicles and bacteria) we used the value $n_{cytoplasm}$ = 1.367 according to the literature (Choi et al., 2007; Marquis, 1973), whereas for the bacterial cell wall we used $n_{cellwall}$ = 1.455 (Marquis, 1973). Finally, we assumed that the dependence of the refractive index as a function of the temperature of the lipid membrane (for vesicles and bacteria) was described by a Boltzmann sigmoidal curve (Equation 1).

# RESULTS AND DISCUSSION

**Isolation of OMVs and MVs**

The overall workflow followed for the isolation of bacterial extracellular vesicles comprehensive of the quality control steps has been depicted in **Figure 1**. Briefly, to have an enriched solution of bacterial vesicles, we started from a large amount of culture medium that has been first centrifuged and filtered to remove bacteria. Then the tangential flow ultrafiltration procedure allowed us to concentrate the sample up to 100 times and reduce the operating time compared to traditional ultrafiltration techniques. The ultrafiltered solution undergone ultracentrifugation to obtain a transparent pellet of bacterial vesicles that after a washing step afforded a very clean sample without residual proteins contaminants. Our extraction and purification protocol allowed us to obtain pure and concentrated samples of OMVs from cultures of *E. coli* grown at three different temperatures (i.e., 20°C, 27°C and 37°C). Similar procedures were adopted for the isolation of MVs from LGG.

**Morphological characterization**

DLS intensity weighed distribution (**Figure 2**) and TEM analysis of OMVs from *E. coli* grown at 37°C (inset of **Figure 2**) revealed that vesicles have a spherical shape with a mean hydrodynamic diameter of 48±3 nm (FWHM=24±2 nm). Notably, the DLS analysis revealed that the vesicle size decreases as the growth temperature is lowered (**Table 1**). OMVs grown at 27°C and 20° C have a mean diameter of 37±4 nm (FWHM=32±2 nm) and 24±2 nm (FWHM=20±3 nm), respectively. Typical TEM images and DLS intensity weighed distributions for vesicles growth at 27°C and 20°C are reported in Supporting Materials.

It is known that *E. coli* can undergo remarkable modifications (i.e., conformational and/or compositional changes of the lipid components) of the outer membrane as a function of the different environmental conditions such as the growth temperature. Therefore, to investigate if the size reduction of OMVs as a function of the growing temperature could be related to a different conformational state of the lipids that form the vesicle itself or their remodeling within the bilayer, we performed turbidimetric measurements on OMV samples extracted from bacteria grown at the same temperatures.

### Turbidimetric experiments

The intensity of light scattered at a fixed angle by the particles in a suspension depends on their size, geometry and optical properties (structure and refractive index of the components). In particular, the refractive index of a lipid bilayer depends on the conformational state of the lipids and on their ordering in the bilayer. Hence, by measuring the time-averaged intensity of scattered light as a function of the temperature, the instrument allowed us to determine the thermotropic lipid phase transitions (with characteristic transition temperatures Tc), related to the lipid organization and their remodeling within the bilayer (Michel et al., 2006; Aleandri et al., 2012). We used the term 'turbidimetric measurements' for this unconventional way of using a light scattering apparatus although this term is not strictly appropriate, since in the present case to reveal possible changes in the refractive index of the lipid membranes as a function of temperature, the scattered light and not the transmitted light, as in conventional turbidimetric measurements, is monitored. Turbidimetric measurements were performed in the temperature range 10°C-45°C on OMVs samples and on bacteria grown at 37°C, 27°C and 20°C.

*Characterization of OMVs extracted from E. coli cultured at 37°C*

In **Figure 3** the results obtained from turbidimetric measurements for vesicles (**Figure 3a**) and *E. coli* (**Figure 3b**) (as comparison) grown at 37°C have been reported. As far as the vesicles are concerned, the mean scattering intensity (**Figure 3a**, upper panel) decreased significantly (about 25%) when the temperature was increased. We determined a transition temperature by fitting a Boltzmann sigmoidal curve [**Equation 1**] to the experimental data (red line in **Figure 3a**, upper panel). The fitting procedure provided a transition temperature of 32 ± 2°C, where the uncertainty is the fitted slope ΔT.

The observed change of the scattered intensity could be ascribed to different factors (Berne and Pecora, 1976) (i.e., to a decrease of the concentration and/or the size of the suspended particles) or to a variation of vesicles refractive index (Michel et al., 2006). However, we did not observe any phase separation (flocculation or precipitation of the particles) and the hydrodynamic diameter of the particles determined by DLS (lower panel of **Figure 3a**) did not show any appreciable dependence on temperature. Hence, we ascribed the observed decrease of the scattered intensity to a change of the refractive index of the particles, likely due to a phase transition of the membrane lipid bilayer or, more in general, to a structural reorganization of the membrane (Berne and Pecora, 1976). **Figure 3b** shows the results obtained from a suspension of *E. coli*. Again, also with bacteria a rather steep variation of the scattered intensity was observed within the same temperature range observed in the case of OMVs suggesting the occurrence of a similar membrane transition. However, in the case of *E. coli* the scattered intensity increased, whereas in the case of OMVs decreased. Although the scattered intensity measurements were unreliable above 35°C (owing to the proliferation of bacteria that altered the optical properties of the suspension), a transition temperature of 30 ± 3°C was obtained (**Figure 3b**). We attributed the opposite behavior of the scattered intensity to the different structures of the bacteria cell wall and, as a consequence, of the OMVs membrane.

*Characterization of OMVs extracted from E. coli cultured at 27°C and 20°C*

Similar results were obtained also for samples grown at 27°C and 20°C. At these growth temperatures, OMVs were smaller (diameters were 37±4 nm and 24±2 nm, respectively, compared to 48±3 nm at 37°C) and their concentration was found to be systematically lower than that obtained for the samples grown at 37°C. As a consequence, also the mean scattering intensity was significantly lower. Nevertheless, also in these conditions the trend of the registered data points suggested the presence of a phase transition. In **Figure 4** we compared the scattering intensity of vesicles (**Figure 4a**) and bacteria (**Figure 4b**) grown at different temperatures. For each sample we reported the normalized scattering intensity, the fitting curve (solid line) and the value of the fitted transition temperature. All the samples analyzed shown a similar behavior in terms of phase transitions. Interestingly, for both OMVs and bacteria a strong correlation between the growth temperature at which *E. coli* was cultured and the transition temperature of the membrane was found (**Figure 5**).

From previous *E. coli* studies, we know that by reducing the temperature of the growth medium the content of unsaturated lipids, which confer a higher flexibility to the membrane, increases consequently. Moreover, bacteria membranes show thermotropic phase transitions whose characteristic transition temperature depends on the lipid composition (Trauble and Overath, 1973) which in turn is related to the growth temperature (Morein et al., 1996). It is reasonable to think that OMVs, that originate from the bacterial outer membrane, may have a similar composition and similar thermotropic behaviors.

**Simulations of bacteria and OMVs scattering cross section**

To better understand the different optical response observed for vesicles and bacteria, we calculated the scattering cross section of different structures simulating vesicles and bacteria by using the Mie theory (Laven, 2003). Vesicles and bacteria have been represented and modeled as shown in **Figure 6**. In particular, OMVs have been modeled as a single-shelled sphere, where the core is the cytoplasm (assumed homogeneous) and the shell represents the lipid membrane, whereas in the case of *E. coli* also a second (internal) shell is added, which represents the cell wall, mainly composed of peptidoglycans. To take into account the thermotropic behavior and evaluate the refractive index n(T) of the lipid bilayer, we assumed the Boltzmann function dependence on the temperature (see **Equation 1)** with the $T_0$ and $\Delta T$ values determined from experimental turbidimetric measurements. We also imposed the ratio of the asymptotic values of the refractive index ($n_0/n_1$) equal to the ratio of the measured intensities ($I_0/I_1$). Finally, we constrained the Boltzmann curve to assume the value of 1.44 at T=23 °C according to the literature (White et al., 2000).

The results of the calculations obtained using MiePlot 4.6 simulator were reported in **Figure 7**. In particular, **Figure 7a** showed the transitions of OMV models, whereas **Figure 7b** the corresponding curves calculated for the *E. coli* models. Notably, the simulations clearly confirmed not only the temperature-dependent transition trend but also the opposite scattering intensity behavior of OMVs and bacteria, as we previously observed experimentally. Simulations gave us the possibility to identify the reason behind the transition differences observed in the case of OMVs and bacteria and ascribe this behavior to the presence of a second 'layer' beneath the outer membrane (the cell wall comprising the peptidoglycan layer) with a fixed refractive index. In fact, this is the only difference between these two models (at least from the optical point of view).

This strongly support the concept the trend of the transition temperature may represent a characteristic 'fingerprint' of OMVs and this behaviour could be employed as a characteristic 'marker' to distinguish vesicles from the bacteria that originated them.

**Small Angle X-ray Scattering experiments**

To investigate in more details the effects of the temperature on OMVs structure we performed SAXS experiments. Due to the low concentration of vesicles obtained at 27°C and 20°C, we performed these measurements only on vesicles grown at 37°C. We reported SAXS patterns in **Figure 8a.** The overall SAXS pattern, which is typical of shelled spheres, did not change significantly when the temperature increased. These data confirmed that OMVs size did not vary in the temperature range investigated and no membrane damages were reported. However, the smooth peak centered at $Q \approx 1$ nm$^{-1}$ (inset of **Figure 8a**) is related to the thickness of the membrane and has been already detected by SAXS on Alix-positive exosome-like small extracellular vesicles (Romancino et al., 2018). This peak significantly changed the shape and moved towards higher Q values at increasing temperatures, although this variation was modest. In order to better understand the significance of this peak position shift, SAXS curves were analysed considering the simplified model of OMVs reported in **Figure 6.** Although OMVs membrane contain a mixture of different lipids (i.e., sterols, polyisoprenoids, etc.) (Bramkamp and Lopez, 2015) and proteins, this simple model appeared to be able to reproduce reasonably our SAXS data. The thickness as a function of temperature obtained from the fitting procedure was reported in **Figure 8b**. The radius and the electron densities as a function of temperature were reported in the Supporting Material. The thickness of the OMVs membrane changed significantly as the temperature increased, whereas vesicles size was not affected by temperature, remained intact in this temperature range as indicated by the core electron density as well as the shell's one that remained constant. These findings suggested that temperature was able to induce a general rearrangement of the membrane structural motifs and that these motifs are linked to the reorganization of the lipid membrane and/or to the conformational changes of its components.

**OMVs have a distinct phase transition behaviour compared to other MVs**

In order to assess the feasibility of using the phase transition profile to discriminate extracellular vesicles coming from different bacteria (i.e., Gram-negative and Gram-positive bacteria) and potentially use this information for diagnostic purposes in the microbiology field, we decided to measure the phase transition properties of MVs from the Gram-positive LGG comparing the results with those obtained with OMVs from the Gram-negative *E. coli*. In this specific case, we studied only the phase transition of MVs from LGG by turbidimetric measurements as this technique is easy to reproduce, does not require expensive equipment, is cheap and also relatively fast. All of these characteristics are highly desirable when thinking to techniques for diagnostics applications.

In **Figure 9** we reported the results of the scattering intensity measurements in the temperature range between 10°C and 45 °C. By examining the profile of MVs from LGG, we observed the presence of multiple transitions, two of them at a T<35°C and another one likely around 37°C. DLS analysis confirmed that the diameter remained constant (as show in Supporting Material) throughout all the temperature range. Similarly to what observed for OMVs, flocculation or precipitation of MVs did not occur, confirming that the transitions are related only to membrane modifications.

The intensity data were fitted with a double Boltzmann function in the region at T<35°C, providing two transition temperatures of 16±2°C and 28±3°C. Interestingly, the scattering profile of MVs was completely different than that of OMVs, suggesting that MVs and OMVs have a different lipid composition of the membrane.

# CONCLUSIONS

In this work the thermotropic behavior of OMVs and bacteria grown at different temperatures have been studied. The analysis of mean scattering intensity as a function of the temperature has shown the presence of peculiar membrane phase transitions characteristic of both OMVs and bacteria. In particular, the transition temperature that we measured allowed to discriminate not only vesicles of different composition (i.e., OMVs and MVs) but also to correlate them to bacteria that originated them and to their culture conditions (i.e., temperature).

The results we obtained have shown that it is possible to distinguish vesicles coming from different bacteria (i.e., gram-positive and gram-negative bacterial) by simply studying the phase transitions of their membranes. In a close future, by implementing this technique, it could be even possible to perform microbiological analysis by discriminating different bacteria in complex mixtures or matrices (i.e., biological fluids, stool, blood, tissues, etc.)

More generally, this technique could be also employed in fields such as oncology, for example to discriminate human exosomes originating from tumors from that produced by normal cells and help clinicians to diagnose different form of tumors.

# ACKNOWLEDGMENTS

Dr. Federica Del Chierico is gratefully acknowledged for her fundamental role in suggesting the best culturing conditions to grow *E. coli* at different temperatures using the M9 minimal medium and for having critically revised and edited the manuscript draft. Prof. Heinz Amenitsch is acknowledged for his precious assistance in SAXS experiments.

probiotics: prevention and management of infections by probiotics. *J. Nutr.* 140, 698S-712S

**Table 1**. Mean hydrodynamic diameter and Full Width Half Maximum (FWHM) of DLS intensity weighed size distribution obtained by CONTIN analysis of correlation function measured at 23°C of OMVs by Escherichia Coli growth at different temperature; reported errors are the standard deviation of three measurements.

| growth temperature (°C) | mean diameter (nm) | FWHM (nm) |
|---|---|---|
| 37 | 48 ± 3 | 24 ± 2 |
| 27 | 37 ± 4 | 32 ± 2 |
| 20 | 24 ± 2 | 20± 3 |

**Figure 1.** Workflow of the procedures followed to isolate and purify bacterial vesicles for downstream measurements.

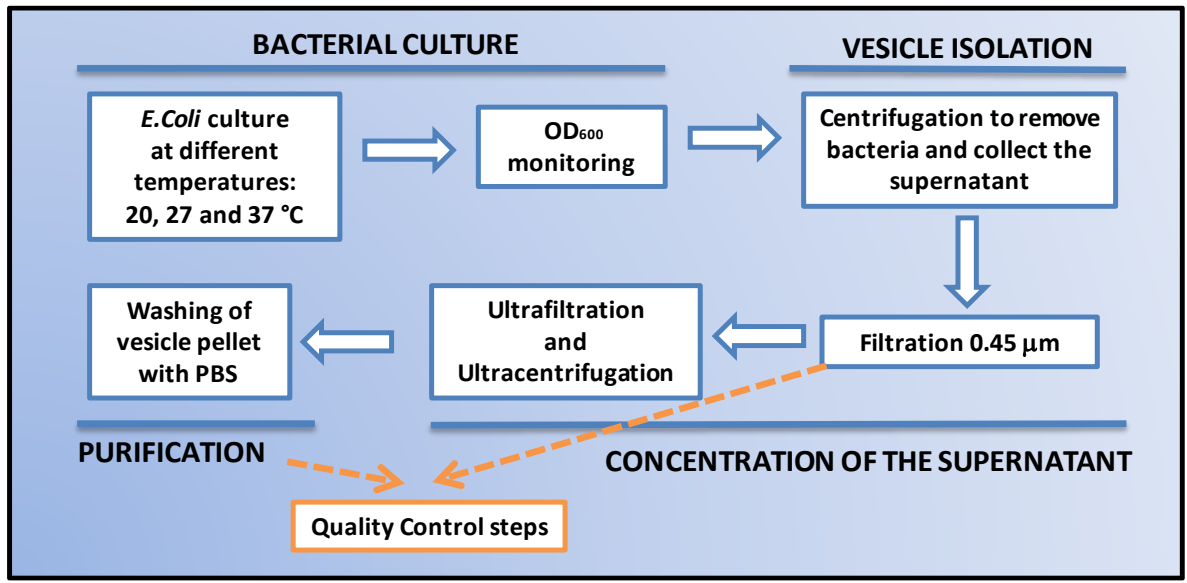

**Figure 2. (a)** Dynamic Light Scattering intensity-weighed distribution of vesicles by *E. coli* grown at 37°C. **(b)** Transmission Electron Microscopy image of vesicles by *E. coli* grown at 37°C obtained without negative staining. In the inset is shown a detail of a TEM image obtained with negative staining.

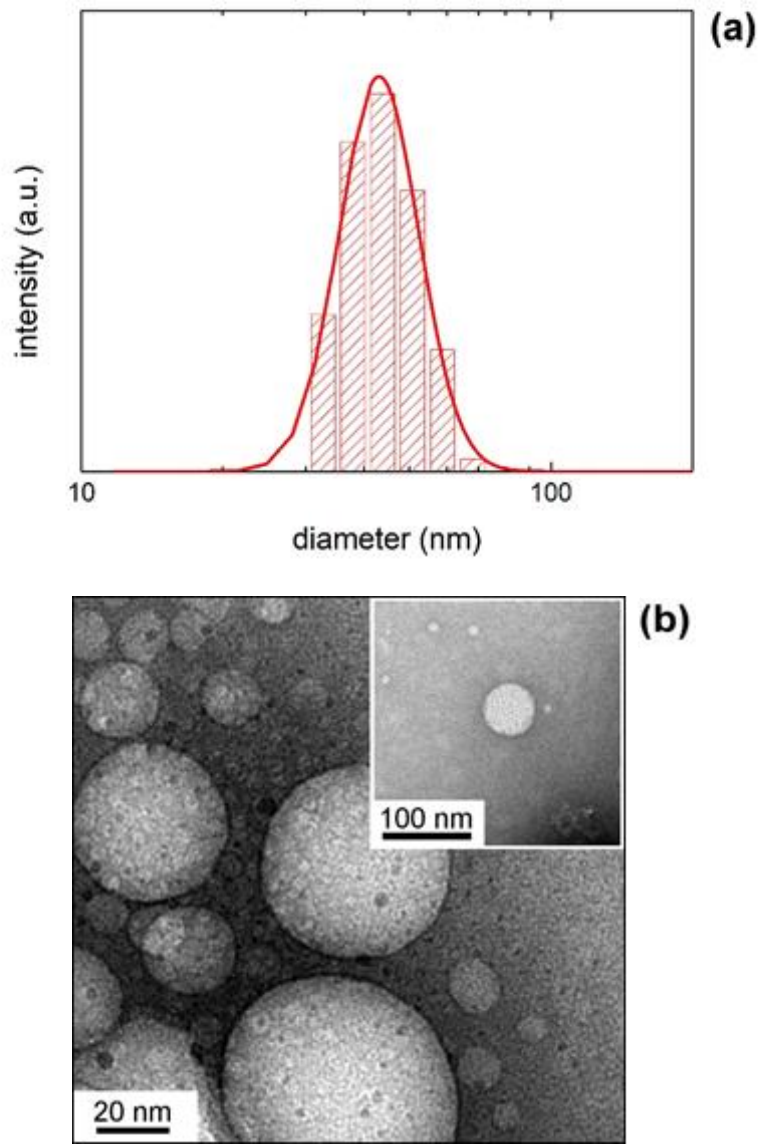

**Figure 3.** (a) Mean scattering intensity (upper panel) and hydrodynamic diameter (lower panel) resulting from DLS measurements at increasing temperatures between 10°C and 45°C of vesicles grown at 37°C; red lines results from Boltzmann fit and linear fit respectively while the blue dashed line highlights the transition temperature. (b) Mean scattering intensity of a DLS measurement at increasing temperatures between 10°C and 35°C of *E. coli* bacteria grown at 37°C. Inset shows the mean scattering intensity up to 45°C (above 35°C the bacteria start to grow).

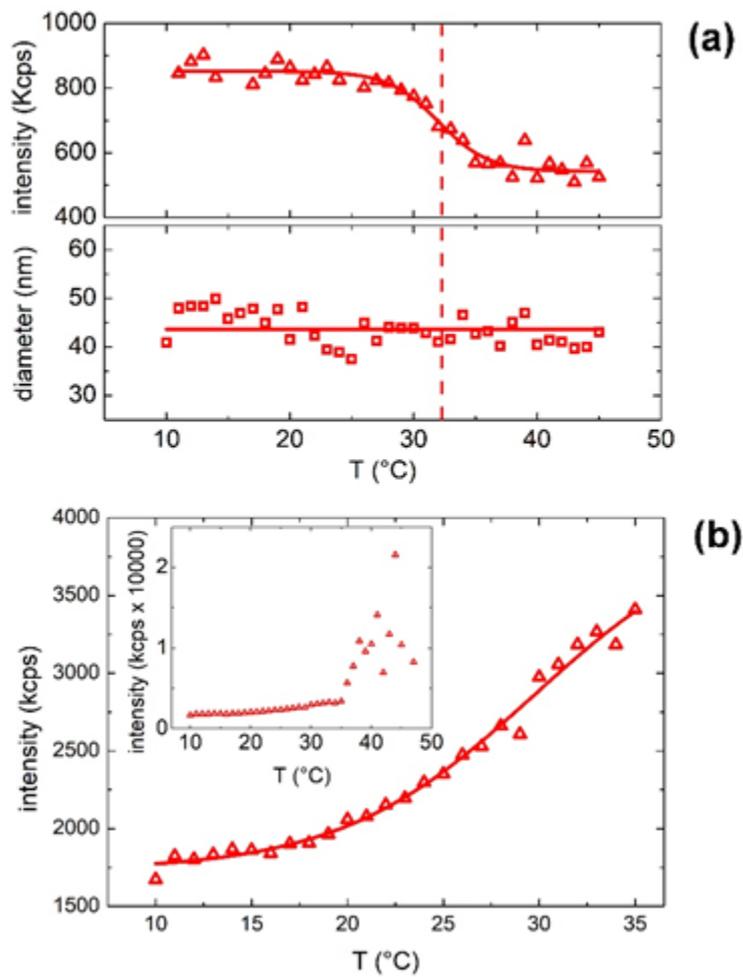

**Figure 4.** Normalized scattering intensity as a function of temperature: comparison between membrane phase transitions of OMVs (a) and E.Coli bacteria (b) grown at 37°C, 27°C and 20°C; vertical dashed lines highlights the transition temperatures.

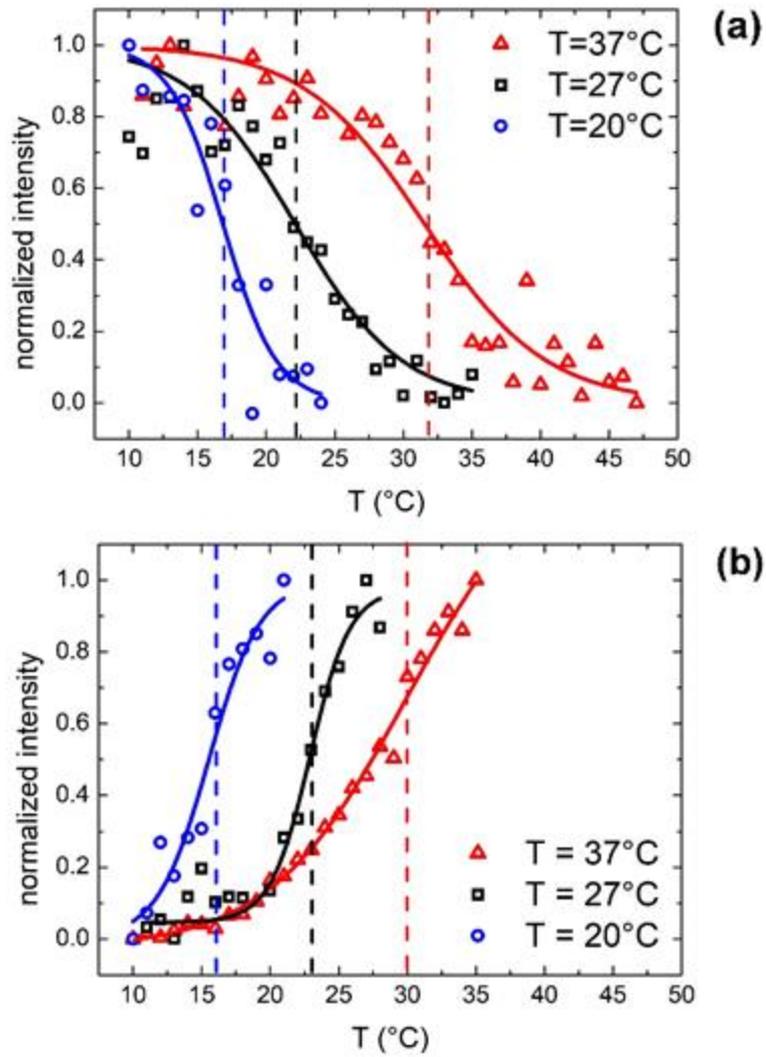

**Figure 5.** Comparison between phase transitions temperature of OMVs and *E. coli* bacteria grown at 37°C, 27°C and 20°C; the error bars represent the transition widths whereas the blue ovals are guide to eyes that highlight the clear separation between transitions from different membranes.

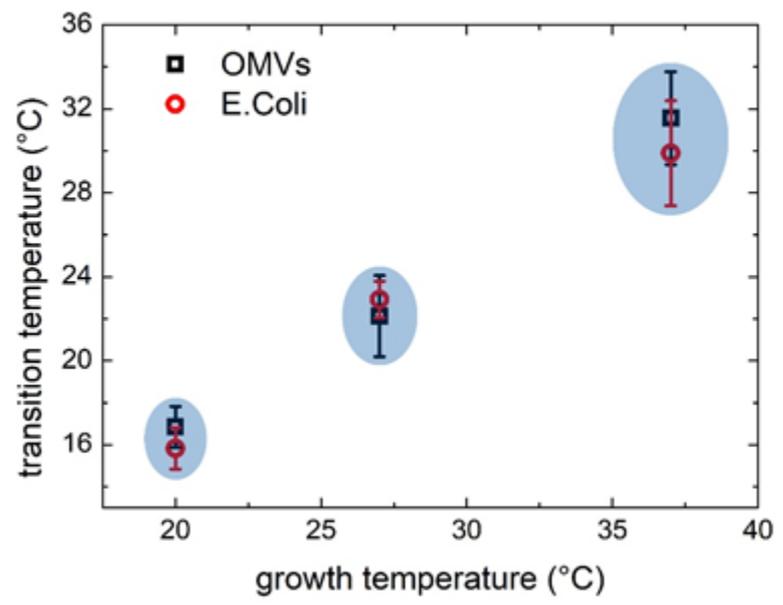

**Figure 6.** Outer membrane vesicle and *E. coli* models representing core and shell structures of different densities (as for SAXS analysis) or different refractive index (scattering intensity simulations).

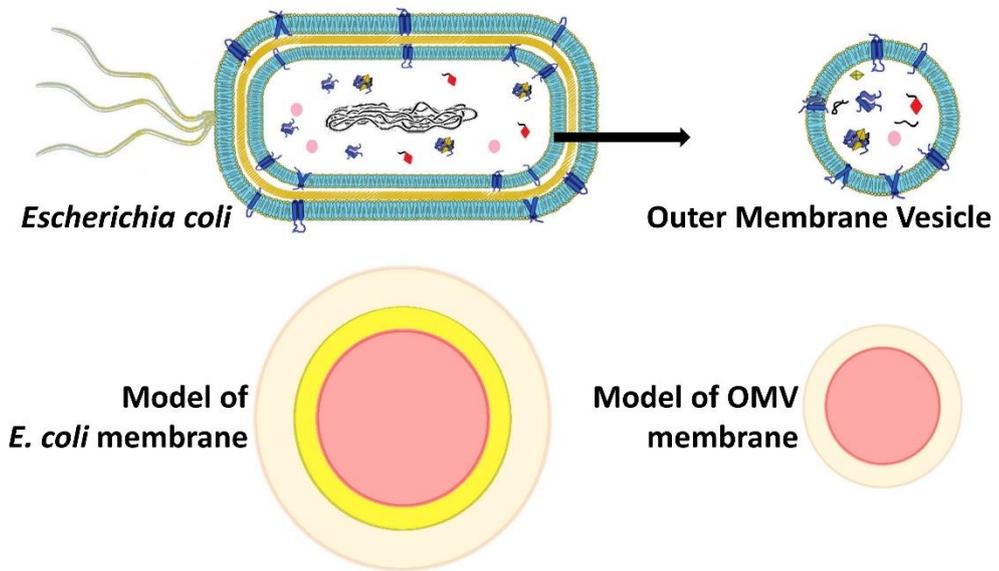

**Figure 7.** Normalized scattered intensity as a function of the temperature: comparison between simulated membrane phase transitions of OMVs (a) and *E. coli* bacteria (b) grown at 37°C, 27°C and 20°C.

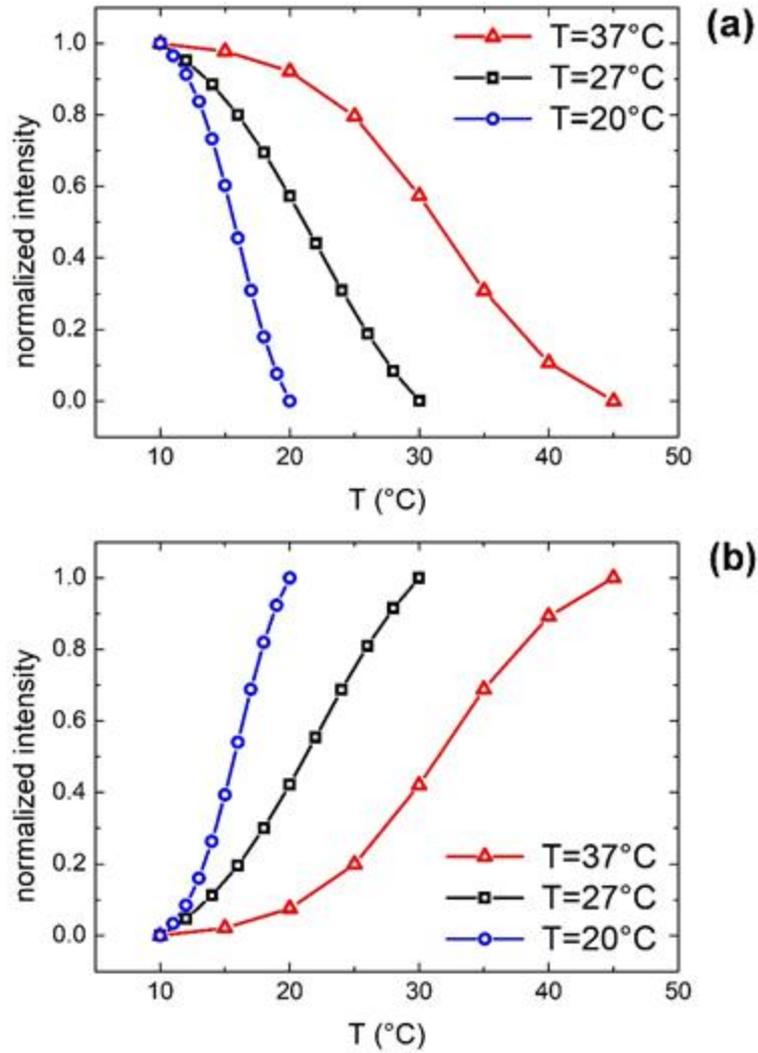

**Figure 8.** (a) SAXS spectra of vesicles maintained at temperatures between 10 and 45°C as in the legend. (b) Vesicles bilayer thickness as resulting from SAXS data fitting procedure, as a function of temperature.

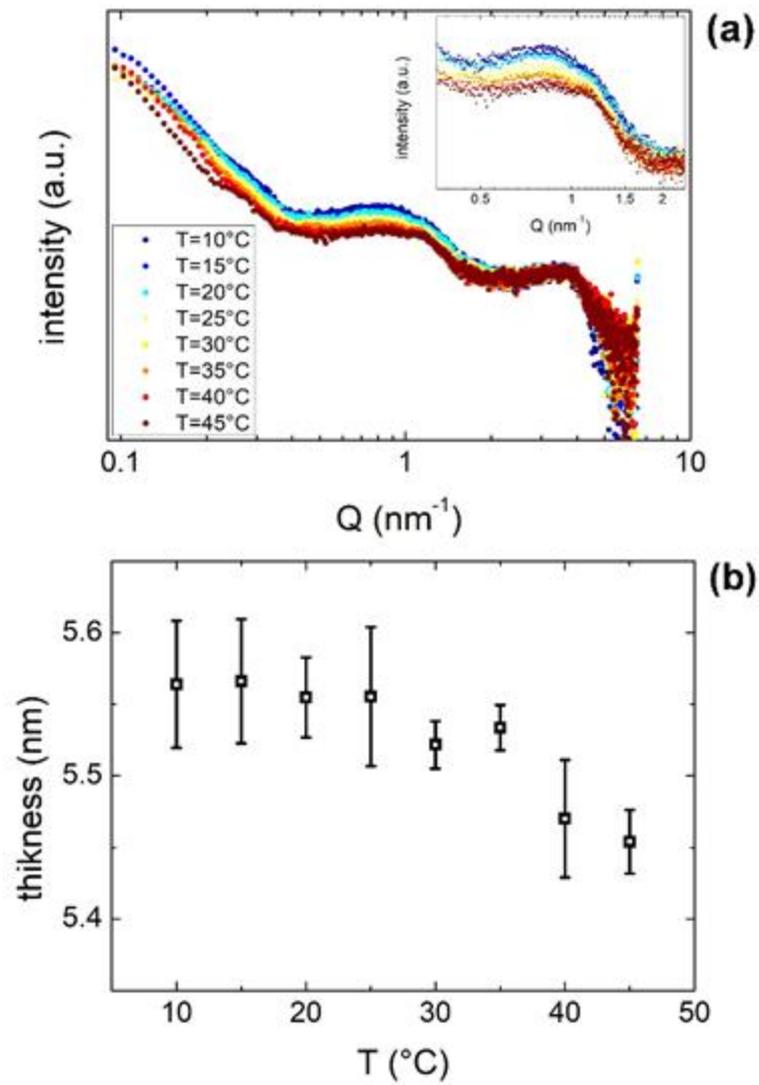

**Figure 9.** Mean scattered intensity of a DLS measurement of MVs by *Lactobacillus rhamnosus LGG* grown at 37°C monitored at temperatures between 10 and 45°C.

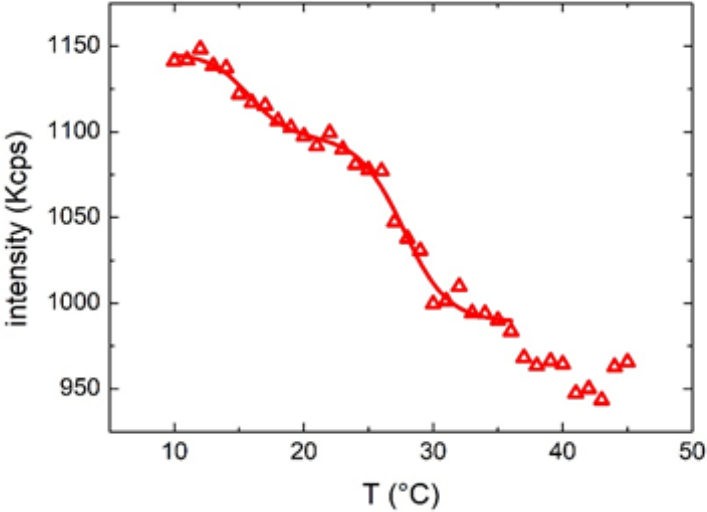